\documentclass[pdflatex,sn-aps,iicol]{sn-jnl}

\usepackage{graphicx}
\usepackage{dcolumn}
\usepackage{bm}
\usepackage[caption=false]{subfig}
\usepackage{amsfonts,amsmath,amssymb,amsthm,braket,nicefrac}
\usepackage{enumitem} 

\usepackage{url}

\usepackage{hyperref}
\hypersetup{
    colorlinks=true,     
    linkcolor=blue,      
    citecolor=blue,      
    filecolor=blue,      
    urlcolor=blue        
}

\usepackage{color}

\usepackage{mathtools} 
\usepackage{xspace} 

\usepackage[capitalise]{cleveref}
\crefname{section}{Sec.}{Secs.}

\newcommand{\SU}{\mathrm{SU}}

\jyear{2022}%

\normalbaroutside 
\begin{document}

\preprint{MIT-CTP/5496}

\title[]{\vspace{-3em} Aspects of scaling and scalability for flow-based sampling of lattice QCD}

\author[1,2]{\fnm{Ryan} \sur{Abbott}}

\author[3]{\fnm{Michael} \spfx{S.} \sur{Albergo}}

\author[6]{\fnm{Aleksandar} \sur{Botev}}

\author[4,1,2]{\fnm{Denis} \sur{Boyda}}

\author[5,3]{\fnm{Kyle}~\sur{Cranmer}}

\author[1,2]{ \fnm{Daniel}~\spfx{C.}~\sur{Hackett}}

\author[6]{\fnm{Alexander~G. D. G.} \sur{Matthews}}

\author[6]{\fnm{S\'{e}bastien}~\sur{Racani\`{e}re}}

\author[6]{\fnm{Ali} \sur{Razavi}}

\author[6]{\fnm{Danilo} \spfx{J.} \sur{Rezende}}

\author[1,2]{\fnm{Fernando} \sur{Romero-L\'opez}}

\author[1,2]{\fnm{Phiala}~\spfx{E.}~\sur{Shanahan}}

\author[1,2]{\fnm{Julian} \spfx{M.} \sur{Urban}}

\affil[1]{\orgdiv{Center for Theoretical Physics}, \orgname{Massachusetts Institute of Technology}, \orgaddress{\city{Cambridge}, \state{MA} \postcode{02139}, \country{USA}}}

\affil[2]{\orgname{The NSF AI Institute for Artificial Intelligence and Fundamental Interactions}}

\affil[3]{\orgdiv{Center for Cosmology and Particle Physics}, \orgname{New York University}, \orgaddress{\city{New York}, \state{NY} \postcode{10003} \country{USA}}}

\affil[4]{\orgdiv{Argonne Leadership Computing Facility}, \orgname{Argonne National Laboratory}, \orgaddress{\city{Lemont}, \state{IL} \postcode{60439} \country{USA}}}

\affil[5]{Physics Department, University of Wisconsin-Madison, Madison, WI 53706, USA}

\affil[6]{\orgname{DeepMind}, \orgaddress{\city{London}, \country{UK}}}

\makeatletter
\def\Authorfont{\reset@font\fontsize{11bp}{14.5bp}\selectfont\boldmath\titraggedcenter}
\def\addressfont{\reset@font\fontsize{9bp}{13.5bp}\selectfont\titraggedcenter}%
\makeatother

\date{November 14, 2022}

\abstract{Recent applications of machine-learned normalizing flows to sampling in lattice field theory suggest that such methods may be able to mitigate critical slowing down and topological freezing. However, these demonstrations have been at the scale of toy models, and it remains to be determined whether they can be applied to state-of-the-art lattice quantum chromodynamics calculations. Assessing the viability of sampling algorithms for lattice field theory at scale has traditionally been accomplished using simple cost scaling laws, but as we discuss in this work, their utility is limited for flow-based approaches. We conclude that flow-based approaches to sampling are better thought of as a broad family of algorithms with different scaling properties, and that scalability must be assessed experimentally.}

\maketitle

\section{Introduction}
\label{sec:intro}

Lattice quantum field theory (LQFT) is the only known systematically improvable approach to calculating physical observables in quantum field theories that exhibit non-perturbative dynamics.
LQFT has been applied to first-principles studies of quantum chromodynamics (QCD) at low energy scales~\cite{morningstar2007,Lehner:2019wvv,Kronfeld:2019nfb,Cirigliano:2019jig,Detmold:2019ghl,Bazavov:2019lgz,Joo:2019byq}, to test proposed models for physics beyond the Standard Model \cite{Brower:2019oor,DeGrand:2015zxa,Svetitsky:2017xqk,Kribs:2016cew}, and to investigate various condensed matter systems~\cite{Ichinose:2014cba,Mathur2016}.
In this framework, discretized path integrals are evaluated numerically using stochastic Monte Carlo estimators,
\begin{equation}
\braket{\mathcal{O}}_p 
\equiv \int d\phi \, p(\phi) \mathcal{O}(\phi) 
\approx \frac{1}{N} \sum_{i=1}^N \mathcal{O}(\phi_i),
\end{equation}
where $\phi_i$ are samples of the lattice field degrees of freedom drawn from the probability distribution defined by the Euclidean lattice action $S$,
\begin{equation}
    p(\phi) = \frac{1}{Z} e^{-S(\phi)}.
\end{equation}
The partition function $Z$ is typically unknown, but this is not an obstacle to sampling with Markov chain Monte Carlo (MCMC) algorithms.
At present, Hybrid/Hamiltonian Monte Carlo (HMC)~\cite{Duane:1987de,neal1993probabilistic,neal1996bayesian} is the state-of-the-art MCMC algorithm for generating QCD field configurations, but its computational cost grows rapidly as the continuum limit is approached~\cite{Wolff:1989wq,Schaefer:2009xx,Schaefer:2010hu}.

Recent work has explored whether improved configuration generation algorithms can be achieved using machine learning (ML)~\cite{%
LiWang2018NNRG,Albergo:2019eim,Albergo:2021vyo,%
Nicoli:2020evf,Nicoli:2020njz,Hackett:2021idh,DelDebbio:2021qwf,
Kanwar:2020xzo,Boyda:2020hsi,Foreman:2021ixr,Foreman:2021ljl,Foreman:2021rhs,%
Albergo:2022qfi,Finkenrath:2022ogg,Abbott:2022zhs,Abbott:2022hkm,%
Albergo:2021bna,%
Gabrie:2021tlu,deHaan:2021erb,Lawrence:2021izu,Jin:2022bgq,Pawlowski:2022rdn,Gerdes:2022eve,Singha:2022lpi,Matthews:2022sds,Caselle:2022acb,Caselle:2022esc,Wang2017,Huang:2017,song2018anicemc,Tanaka:2017niz,levy2018generalizing,Pawlowski:2018qxs,Cossu:2018pxj,Wu:2019,Bachtis:2020dmf,Nagai:2020jar,Tomiya:2021ywc,Bachtis:2021eww,Wu:2021tfb,Tomiya:2022meu,Mate:2022jtr}.
For example, promising proof-of-principle results using normalizing flows~\cite{rezende2016variational,dinh2017density,papamakarios2019normalizing} have been obtained in a range of theories with different properties including gauge symmetries~\cite{Kanwar:2020xzo,Boyda:2020hsi,Albergo:2022qfi,Abbott:2022zhs,Abbott:2022hkm,Finkenrath:2022ogg,Foreman:2021ljl,Foreman:2021ixr,Foreman:2021rhs}, fermionic degrees of freedom~\cite{Albergo:2021bna,Finkenrath:2022ogg,Albergo:2022qfi,Abbott:2022zhs,Abbott:2022hkm}, and the existence of distinct topological sectors or multiple modes of the probability distribution~\cite{Kanwar:2020xzo,Hackett:2021idh,Albergo:2022qfi,Nicoli:2020evf,Nicoli:2020njz,Finkenrath:2022ogg,Foreman:2021ljl,Foreman:2021ixr,Foreman:2021rhs,Abbott:2022zhs,Abbott:2022hkm,DelDebbio:2021qwf}.
This includes a first demonstration for lattice QCD~\cite{Abbott:2022hkm}.

These results raise the question of whether a flow-based approach can be applied to lattice QCD calculations at state-of-the-art scale. The practical question is whether flow models can provide more cost-effective sampling than HMC for QCD at parameters and volumes of interest. Discussions of this question often conflate two separate concerns,
\setlist{nolistsep}
\begin{itemize}[itemsep=1ex, leftmargin=*, topsep=1ex]
\item \textit{Scalability:} whether an approach can be practically applied to some target theory, for which the ultimate question is the \emph{cost} to generate $N$ samples at a given set of parameters;
\item \textit{Cost scaling:} how the cost of an approach changes as certain parameters are varied---in this case, the parameters of the target lattice QCD theory, including the parameters of the action and the lattice geometry.
\end{itemize}
When we ask whether an approach to gauge field sampling for QCD is ``scalable,'' we are asking not about its precise scaling properties, but whether it will work for QCD on state-of-the-art volumes at state-of-the-art parameters, and whether it can be used to push the state of the art further.
For the sampling of gauge field configurations using HMC, cost scaling relations often directly and usefully predict scalability.
However, the range of applicability, and hence utility, of any practical cost-scaling relations for flow-based algorithms is much more limited.
The rest of this manuscript is devoted to elucidating this statement.

To that end, \cref{sec:prelims} first reviews HMC and flow-based sampling methods for lattice field theory, establishing definitions and notation for the rest of the discussion.
\Cref{sec:flows-vs-hmc} then compares flow-based approaches with HMC, emphasizing counterintuitive differences to the more familiar HMC paradigm.
\Cref{sec:flow-demos} presents numerical illustrations of different aspects of the scaling of flow-based approaches, demonstrating the limitations of scaling laws in assessing scalability.
Finally, \cref{sec:outlook} speculates on potential paradigms for flow-based sampling at scale and discusses the outlook for these methods.

\section{Preliminaries}
\label{sec:prelims}

\subsection{HMC}
\label{sec:hmc-review}

In the HMC approach, gauge field configurations are generated in a Markov chain, where approximate molecular dynamics evolution in the fictitious ``Monte Carlo time'' direction is used to propose updates to the chain.
Each step of the HMC algorithm proceeds by drawing fictitious momentum variables conjugate to each lattice field degree of freedom, integrating the Hamiltonian equations of motion, then accepting or rejecting the resulting new field configuration with a Metropolis test to correct for integrator errors and guarantee exactness of sampling~\cite{Duane:1987de,neal1993probabilistic,neal1996bayesian}.
For gauge theories with fermion content, such as QCD, the gauge fields are evolved in a fixed background of auxiliary ``pseudofermion'' fields, which encode fermionic effects stochastically~\cite{Weingarten:1980hx,Fucito:1980fh}.
In this case, each integrator step requires solving large sparse systems of linear equations using implicit methods like conjugate gradient.
For QCD, these solves typically dominate the computational cost.
See e.g.~Refs.~\cite{neal2011mcmc,Gattringer:2010zz,DeGrand:2006zz} for more detailed reviews.

\subsection{Normalizing flows and flow-based sampling}
\label{sec:flow-review}

Normalizing flows are a framework for building exact, numerically tractable, machine-learned maps between probability distributions.
In this construction, a diffeomorphic flow function $f$ is applied to transform samples $z$ drawn from a base distribution $r$ to obtain samples $\phi = f(z)$ distributed according to a model distribution $q$.
The flow $f$ is parametrized by neural networks and can be optimized to some objective by the minimization of a ``loss function''. 
Conservation of probability gives the density of transformed samples,
\begin{equation}
q(\phi) = \left| \frac{\partial f(z)}{\partial z} \right|^{-1} r(z) ~ .
\end{equation}
Various frameworks have been developed to construct expressive, trainable flow transformations for which this expression can be evaluated tractably~\cite{dinh2014nice,dinh2017density,durkan2019neural,Rezende:2020hrd,Kanwar:2020xzo,Boyda:2020hsi,Albergo:2021bna,Abbott:2022zhs,ZhangEWang2018Monge,huang2020convex,pmlr-v70-amos17b,chen2018neural}.

Flow transformations are a general tool which can be applied 
as components of different sampling approaches (and in non-sampling applications).
In particular, in the ``direct sampling'' approach, which is the primary concern of this work, flows relate a simple, easy-to-sample base distribution to a learned approximation $q \approx p$ of the target lattice field distribution, ${p=\exp[-S]/Z}$.
Evaluating the reweighting factors $w(\phi)=p(\phi)/q(\phi)$ between the model $q$ and target $p$ provides sufficient information to compute expectations under $p$.
This may be accomplished e.g.~by computing reweighted expectations as $\braket{\mathcal{O}}_p = \braket{w \mathcal{O}}_q$.
Alternatively, samples from $p$ can be generated by using flow model samples as proposals for the Independence Metropolis algorithm~\cite{Metropolis:1953am,Hastings:1970aa,tierney1994markov} with acceptance probability ${p_\text{acc}(\phi \rightarrow \phi') = \min[1, w(\phi')/w(\phi)]}$.
Powerfully, this approach allows composition with other MCMC algorithms with complementary properties~\cite{Hackett:2021idh}.
As discussed further below, these straightforward applications are only a few examples of how flow-based sampling algorithms may be constructed.

Even within a particular sampling framework, there is no unique way to construct a flow model. Each concrete realization---a model architecture---is defined by many different choices.
From high- to low-level, these are broadly: 
\setlist{nolistsep}
\begin{itemize}[noitemsep, leftmargin=1.5em]
\item domain of the variables $z$, $\phi$ (e.g.~$\mathbb{R}$, $\SU(N)$, multiple fields);
\item the choice of base distribution $r(z)$ (e.g.~Gaussian, Haar uniform, free theory); 
\item strategy for constructing flow functions (e.g.~convex potential flows~\cite{ZhangEWang2018Monge,huang2020convex,pmlr-v70-amos17b}, neural ODEs~\cite{chen2018neural}, or coupling layers~\cite{dinh2014nice,dinh2017density} constructed using affine transformations~\cite{dinh2017density}, non-compact projections~\cite{Rezende:2020hrd}, neural splines~\cite{durkan2019neural}, or gauge-equivariant transformations~\cite{Kanwar:2020xzo,Boyda:2020hsi,Abbott:2022zhs});
\item structure of neural networks parametrizing the flow transformation (e.g.~fully connected vs.~convolutional networks, choice of activation functions);
\item and various hyperparameters (e.g.~number of coupling layers, neural network depths and widths).
\end{itemize}

While flow-based sampling approaches in principle guarantee unbiased results by construction even for models with $q$ arbitrarily different from $p$, performant sampling requires training the models.
Just as for model architectures, many choices define a particular training scheme, including:
\setlist{nolistsep}
\begin{itemize}[noitemsep, leftmargin=1.5em]
\item scheme for initializing model parameters (e.g.~random distribution, retraining);
\item approach to training data (e.g.~self-training, training on existing configurations);
\item choice of loss function to minimize (e.g.~KL divergence~\cite{Kullback:1951}, Stein discrepancies~\cite{liu2016kernelized,gorham2017measuring}, or gradient-based divergences like score matching or Fisher divergence~\cite{hyvarinen2005estimation,johnson2004information});
\item optimization algorithm (e.g.~SGD, Adam~\cite{kingma2017adam}, second-order optimizers);
\item all hyperparameters of training (e.g.~optimizer parameters, batch size);
\item and the schedule for training, which may vary any or all of these choices over time.
\end{itemize}

Besides direct sampling, other approaches to sampling lattice field distributions have been explored which use normalizing flows for different statistical modeling tasks.
For example, Ref.~\cite{Finkenrath:2022ogg} explored the use of flows which model a localized patch of a lattice field, conditioned on its environment.
Refs.~\cite{Foreman:2021ixr,Foreman:2021rhs} employed flows to generalize the proposal distribution in HMC.
Ref.~\cite{Albergo:2021bna} used flows to model conditional distributions for Gibbs sampling.
Stochastic normalizing flows~\cite{wu2020stochastic,nielsen2020survae} and related approaches~\cite{dibak2020temperature,arbel2021annealed,Matthews:2022sds} use flows to relate a sequence of distributions interpolating between the base and target distributions.
Flows play a different role in each case, requiring different architectures and training schemes.

\subsection{Cost decomposition}

For a generic sampling algorithm, the cost to generate a dataset equivalent to $N_\text{indep}$ independent samples from a target distribution can be decomposed as
\begin{equation}
    C_\text{total}(N_\text{indep}) = C_\text{setup} + C_\text{samp}(N_\text{indep}) ~ ,
    \label{eq:total-cost-decomp}
\end{equation}
where $C_\text{setup}$ is any up-front cost that must be paid before beginning data generation (e.g.~the cost of training for flows, or of equilibration or decorrelating a forked stream for HMC), and $C_\text{samp}$ is the cost of sampling thereafter.
Generically, sampling costs may be further decomposed as
\begin{equation}\begin{aligned}
    C_\text{samp}(N_\text{indep})
    &= \Delta C_\text{samp} \, N_\text{raw}(N_\text{indep})
    \\
    &= \Delta C_\text{samp} \frac{ N_\text{indep}}{ \text{ESS}} ~ ,
    \label{eq:samp-cost-decomp}
\end{aligned}\end{equation}
where $\Delta C_\text{samp}$ is the cost to generate a single configuration (e.g.~the cost of a single HMC trajectory, or of drawing a sample from a flow model) and $N_\text{raw}$ is the total number of configurations output by the procedure.
The effective sample size per configuration ${\text{ESS} \equiv N_\text{indep} / N_\text{raw} \in [0,1]}$ quantifies the loss of statistical power due to e.g.~MCMC autocorrelations or the increase in variance due to reweighting.
For HMC, $\text{ESS} = 1 / 2 \tau_\text{int}$, where $\tau_\text{int}$ is the integrated autocorrelation time. For flow-based direct sampling, the reweighting-inspired metric $\text{ESS} = 1/\braket{w^2}_q$ \cite{doucet2001sequential,liu2001monte} is commonly employed (see Appendix~\ref{sec:numerics-appendix} for further discussion).
Note however that no single scalar metric may fully quantify the performance of any sampling algorithm, as the true ESS is always observable-dependent: for HMC, observables are sensitive to different autocorrelation times, whereas for flows, each observable has different correlations with the reweighting factors.
Both $C_\text{setup}$ and $\Delta C_\text{samp}$ carry units of compute time (e.g.~GPU hours), and depend strongly on the precise details of the algorithm implementation and hardware.

\section{Costs and cost scaling: HMC versus flow-based approaches}
\label{sec:flows-vs-hmc}

In this section, we first analyze the role and limitations of cost scaling relations in the familiar context of HMC for QCD.
Subsequently, we discuss different general aspects of flow-based approaches and their cost scaling properties, emphasizing differences with HMC.
Features better demonstrated with numerical examples are deferred to \cref{sec:flow-demos}.

\subsection{Expectations for scaling laws from HMC}

The cost to sample QCD field configurations using HMC is often parametrized as a function of the physical parameters of the target theory as \cite{Gattringer:2010zz,DelDebbio:2015qpg}
\begin{equation}
    C(N_\text{indep}) \propto N_\text{indep} \left( \frac{L}{a} \right)^{z_L} M_\pi^{-z_M} a^{-z_a}
    \label{eqn:hmc-scaling}
\end{equation}
where $a$ is the lattice spacing, $L/a$ is the extent of the lattice in units of sites, and $M_\pi$ is the pion mass.
The exponents $z_L$, $z_M$, and $z_a$ define the cost scaling relation.
A key aspect of the difficulty in assessing scalability is that their values depend on many factors, in particular both algorithmic choices and the targeted regime of physical parameters, i.e.,
\begin{enumerate}[noitemsep]
    \item Scaling laws depend on how algorithm parameters are varied, and
    \item Scaling laws have limited regimes of applicability.
\end{enumerate}
Naturally, not only the parameters of an algorithm are relevant, but also the structure of the algorithm itself, with the important consequence that
\begin{enumerate}[resume]
    \item Algorithmic developments can improve scaling properties.
\end{enumerate}
Each of these key aspects are elaborated on below.
\\

{\noindent \bf  1. Scaling laws depend on how algorithm parameters are varied.}
Defining and quantifying a scaling relation requires choosing some scheme to vary algorithm parameters as the parameters of the target theory are varied.
Different choices will yield different scaling relations. 
For example, discussions of HMC scaling typically  
quote a value of $z_L = 5$ for the exponent of $L/a$ in \cref{eqn:hmc-scaling}.
Part of this, $(L/a)^4$, is due to simple operation counting on a four-dimensional lattice.
However, the remaining factor of $L/a$ follows from choosing a particular scheme to vary the algorithm parameters with the volume, specifically increasing the number of integrator steps per trajectory to keep the acceptance rate fixed~\cite{beskos2013optimal}.
A different choice would yield a different scaling relation: for example, keeping the step size fixed will result in a rapid decline of the acceptance rate, inducing long autocorrelation times and poorer scaling.
\\

{\noindent \bf  2. Scaling laws have limited regimes of applicability.}
Although universal behavior may be expected in some regimes of target physical parameters, in practice a wide range of values of the exponent of the lattice spacing $z_a$ have been quoted in the literature~\cite{Schaefer:2012tq,Gattringer:2010zz}. 
For example, many early studies~\cite{Hasenbusch:2003vg,Ukawa:2002pc,Jegerlehner:2000xt,Lippert:2002jm} of HMC cost scaling found $z_a$ near its conjectured lower bound of 2~\cite{Luscher:2011qa}.
However, approaching the continuum limit, increasingly large potential barriers prevent HMC from tunneling between disjoint topological sectors.
This effect, known as ``topological freezing'', induces very long autocorrelation times and thus large $z_a$.
In fact, the $a$-dependence may instead be consistent with exponential~\cite{Schaefer:2010hu,DelDebbio:2004xh}.
This large variation demonstrates the importance of assessing the limitations of any scaling law: extrapolating using an incorrect value of $z_a$ for the parameter regime under consideration will not be predictive.
\\


{\noindent \bf  3. Algorithmic developments can improve scaling properties.}
Historically, new algorithmic developments have led to qualitative improvements in the capabilities and reach of HMC.
For example, the scaling relation \cref{eqn:hmc-scaling} was infamously used to diagnose the ``Berlin wall'': at the time (ca.~2001), large measured values of $z_M$ implied that calculations at physical pion masses would be practically impossible with near-future hardware~\cite{DelDebbio:2015qpg,Ukawa:2002pc}.
However, Hasenbusch preconditioning~\cite{Hasenbusch:2001ne} was developed soon afterwards, reducing $z_M$ and opening access to the physical pion mass regime.
The development of multigrid preconditioners has provided additional improvements~\cite{Brannick:2007ue,Babich:2010qb}.

\subsection{Costs and cost scaling for flow-based approaches}
\label{sec:general-costs-flow}

History has demonstrated that algorithmic advances can redefine the limits of LQFT methods and enable new physics results.
To that end, flow-based approaches provide an unprecedented new space of algorithms to explore.
However, their costs and cost scaling properties can be counterintuitive from the HMC perspective.
To understand these methods and their scaling properties, several key features of flow models must be appreciated, in particular that:
\setlist{nolistsep}
\begin{enumerate}[noitemsep]
    \item Model weights are algorithm parameters;
    \item Flow evaluation costs do not typically vary with physical parameters, but model quality does.
\end{enumerate}
In addition, the practical requirement for model training introduces conceptual complications. Specifically:
\begin{enumerate}[resume]
    \item Training and sampling are different dynamical processes, but cannot be considered independently;
    \item Training costs may be significant.
\end{enumerate}
Finally, certain low-level computational concerns must be considered:
\begin{enumerate}[resume]
    \item Operation counting can predict raw computational costs; and
    \item Flow-based approaches may parallelize more effectively than serial samplers.
\end{enumerate}
Each of these points is discussed below, emphasizing the conceptual differences between flow-based sampling approaches and the HMC paradigm.
\\

{\noindent \bf 1. Model weights are algorithm parameters.}
A fixed flow model architecture represents a parametric family of probability distributions. Within this family, a set of values for the neural network parameters $\theta$ (``model weights'') specifies a particular model distribution, $q_\theta$.
The ESS to sample some target theory $p$ will vary as a function of \emph{every} model weight; their number thus defines the dimension of the space of algorithms to be explored.
Relaxing the restriction to fixed architectures presents even further possibilities.

Conceptually, this is a drastically different situation from HMC, where the space of algorithm parameters is tractable to explore; flow models may have potentially billions of parameters.
Of course, in practice, hand-tuning is impossible and model weights must instead be set implicitly by training e.g.~with stochastic gradient descent methods.
The practical analog to the set of HMC's algorithm parameters are thus the set of all of the choices involved in defining a training scheme and architecture described in \cref{sec:flow-review}. 
Different scaling behavior results from how these choices are varied with target parameters.
Note that although training schemes are often described in terms of relatively few hyperparameters, this apparent reduction in complexity is artificial. As discussed further below, the need for training only \emph{adds} complexity, rather than providing any simplification.
\\

{\noindent \bf 2. Flow evaluation costs do not typically vary with physical parameters, but model quality does.}
For HMC, the cost to generate each sample, $\Delta C_\text{samp}$, varies strongly with the parameters of the target theory due to the changing problem difficulty for the linear solvers.
In contrast, applying a flow transformation typically involves only explicit algebraic operations.
In this case, $\Delta C_\text{samp}$ is independent of the precise values of the weights of the model, and thus the target parameters they implicitly encode.
It follows that, for a fixed architecture, the dominant cost scaling with target parameters is due to the variation of the ESS (note that this may not hold if $\Delta C_\text{samp}$ is dominated by linear solves to evaluate $p$, or e.g.~for hybrid algorithms incorporating HMC updates).
This severely limits the applicability of empirically measured cost scaling relations: as explored below, model quality is an unpredictable consequence of the complicated interplay between architecture, training scheme, and target theory.
\\

{\noindent \bf 3. Training and sampling are different dynamical processes, but cannot be considered independently.}
\label{sec:samp-train-different}\label{sec:stop-conditions}
For HMC, the same process---evolution by Hamiltonian dynamics---is used for both setup and sampling.
However, the same is not true for flows.
Training is a search process in the high-dimensional space of model parameters, typically involving stochastic gradient descent, with dynamics arising from all the choices described in \cref{sec:flow-review}.
Sampling dynamics depend on the interaction of fine-grained details of model quality with the role played by the model in the sampling algorithm (giving rise to e.g.~the distribution of rejection run lengths in direct sampling with Metropolis).
Unlike for HMC sampling, there is no reason to expect any common parametrization to apply to both cost components.

However, this does not mean that scaling behaviors of training and sampling costs can be studied in isolation.
For flows, sampling efficiency is a function of expenditure on training, i.e., $\text{ESS} = \text{ESS}(C_\text{train})$.
Among other consequences, this implies that the scaling behavior of sampling and training costs are not defined without precisely specifying the stopping condition for training.
Different choices of stopping condition result in qualitatively different scaling relations.
For example, if training always uses a fixed amount of computation, then $C_\text{train}$ scales trivially with the target parameters by construction, and all cost scaling is pushed on to $C_\text{samp}$, as explored in \cref{sec:model-cost-scaling}.
Instead, one may choose to train until some target ESS is achieved.
In this case, the ESS scales trivially while $C_\text{train}$ alone varies, as explored in \cref{sec:training-cost-scaling}.
As demonstrated in \cref{sec:training-cost-scaling,sec:model-cost-scaling}, even the precise choice of fixed $C_\text{train}$ or target ESS can strongly affect scaling properties.
Other choices of stopping condition between these extremes are possible, each inducing different scaling properties.
This sensitivity implies that scaling behaviors for flows are highly non-generic, thus the generalizability of empirically assessed scaling relations is severely limited.
\\

{\noindent \bf 4. Training costs may be significant.}
Most existing applications of flows to sampling in LQFT have employed training methods which apply the flow (or its inverse) to batches of field samples.
For such a training approach, training costs can be decomposed similarly to the decomposition of sampling costs in~\cref{eq:samp-cost-decomp}:
\begin{equation}
    C_\text{setup} = \Delta C_\text{train} N_\text{train}
\end{equation}
where $N_\text{train}$ is the total number of samples flowed during training and $\Delta C_\text{train}$ is the combined cost of flowing a single sample and then backpropagating gradients.
Typically, $\Delta C_\text{train}$ is a few times larger than $\Delta C_\text{samp}$.
Training from a random initialization often involves $\sim$ 1000s of optimizer steps using batches of $\sim$ 100s to 1000s of configurations, meaning $N_\text{train}$ can be much larger than typical QCD ensemble sizes.
This potentially large scale for training costs is an important factor in determining what will be computationally viable for QCD.
However, efficient parallelization (as discussed below) may help mitigate this cost in practice.
Moreover, importantly, as explored in \cref{sec:flow-demos}, these costs are highly non-generic and can be optimized significantly.
\\

{\noindent \bf 5. Operation counting can predict raw computational costs.}
In some cases, simple operation counting can yield relations describing the dependence of  $\Delta C_\text{samp}$ and $\Delta C_\text{train}$ ($\Delta C$, collectively) on model hyperparameter choices and certain target theory parameters, such as the number of lattice sites $\Omega$.
For example, for flow models built from composed sub-transformations (e.g., coupling layers), the cost of applying a flow is linear in the number of sub-transformations. 
Similarly, for flow transformations parametrized by neural networks which interconnect all lattice sites, operation counting predicts that $\Delta C \propto \Omega^2$. 
More favorably, transformations parametrized by convolutions over the lattice geometry may have ${\Delta C \propto \Omega}$.
(Note that these relations are often broken by e.g.~cache effects or adaptive algorithm swapping, and may not apply for implementations on hardware such as GPUs and TPUs if their parallelism is not fully utilized.)
Operation counting thus provides a useful but limited tool for assessing scaling properties and scalability: if the architecture itself is varied with the target, operation counting can predict scaling of $\Delta C$, but cannot account for the (equally important) effect of varying model quality.
Real-world resource constraints may induce further complications, e.g.~if limited memory necessitates a trade-off between model size and the batch size for training.
\\

{\noindent \bf 6. Flow-based approaches may parallelize more effectively than serial samplers.}
Sampling from a flow model is embarrassingly parallelizable, and thus flow-based sampling admits new strategies for parallelism unavailable in the context of serial MCMC algorithms like HMC.
For example, multiple compute nodes might independently and locally generate and store configurations $\phi_i$, and pass only $q(\phi_i)$ and $p(\phi_i)$ to a central coordinator which updates the state of a global, distributed Markov chain with Independence Metropolis.
Potential new strategies include not only the running of multiple independent samplers (without incurring additional setup costs), but also new schemes for problem division such as pipeline parallelism.
Separately, the high arithmetic density of neural network operations suggests that flow-based sampling may not be communications-bound.
Taken together, these properties suggest potential for flow-based methods to make more efficient use of computational resources than serial sampling algorithms, given real-world walltime and hardware constraints.

\section{Limitations of scaling relations for flow-based approaches}
\label{sec:flow-demos}

The previous section discussed properties of flow-based sampling approaches for LQFT that are apparent from the definition of the approach.
However, research and experimentation in this area has identified other properties of flows relevant to assessing scalability.
This section presents simple numerical demonstrations of these features and discusses their implications for the utility of scaling relations.
Further details of all numerical experiments are provided in Appendix~\ref{sec:numerics-appendix}.

\subsection{Training costs}

As discussed in \cref{sec:general-costs-flow}, the scale of training costs of flow models can be significant, requiring the flow to be applied to many more configurations than typical ensemble sizes.
However, as demonstrated here, these costs are highly non-generic, and may be optimized significantly.
Furthermore, their scaling behavior depends on the precise details of the training protocol.
Specifically, we demonstrate that:
\setlist{nolistsep}
\begin{enumerate}[noitemsep]
    \item Training costs may be optimized by orders of magnitude,
    \item Transfer learning can mitigate training costs, and
    \item Training cost scaling depends on training protocol.
\end{enumerate}
The necessary conclusion is that the scaling of training costs is entirely specific to the approach employed.
\\

\begin{figure}[t!]
    \includegraphics[width=1\linewidth]{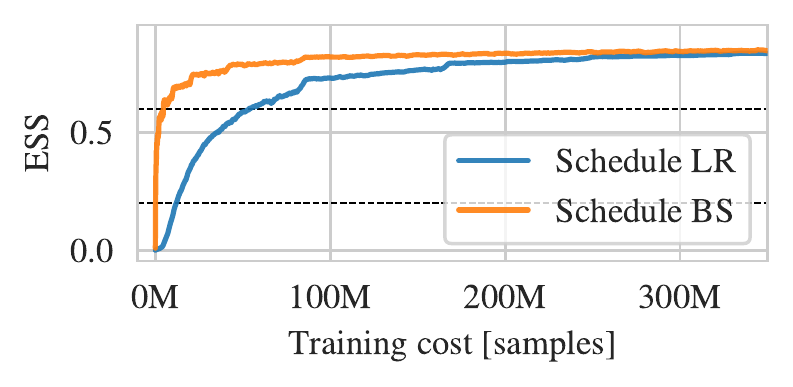}
    \caption{
        Model quality as a function of training cost for two approaches to training, learning rate (LR) scheduling and batch size (BS) scheduling, for a flow model targeting real scalar field theory for $m^2=-1$ and $\lambda=1$ on a $10 \times 10$ lattice.
        Each training scheme is applied to the same architecture.
        Cost is measured in units of the number of samples generated for training, $N_\text{train}$.
        Dashed black lines indicate potential target ESSes of 0.2 and 0.6, as referenced in the main text.
        The batch size for ``Schedule LR'' is 16384 throughout, while for ``Schedule BS'' it begins at 128 and doubles every 5000 steps until it reaches 16384.
        For ``Schedule LR'', the learning rate is halved every 5000 steps.
        The ESS is measured using the training batch size every step, and smoothed over a rolling window of width 250 steps.
        Results vary by $\sim 5\%$ under repeated experiments with different pseudorandom seeds.
        Further details are provided in Appendix~\ref{sec:numerics-appendix}.
    }
    \label{fig:phi4_bs_vs_lr_sched}
\end{figure}

{\noindent \bf 1. Training costs may be optimized by orders of magnitude.}
Periodically reducing the optimization step-size (i.e., the learning rate)---a standard ML technique---can produce models of better quality at lower computational cost than training with a fixed learning rate.
An alternative is to periodically increase the batch size during training~\cite{smith2017don}.
As demonstrated in \cref{fig:phi4_bs_vs_lr_sched}, in some cases this simple change can produce models of equivalent quality for drastically reduced cost as compared with those produced using learning rate scheduling.
The precise improvement factor depends on the criterion used to stop training.
For example, in this demonstration, batch size scheduling is $\sim 110$ times less expensive than learning rate scheduling for training to a target ESS of 0.2, while for a target ESS of 0.6 the improvement is only a factor of $\sim 12$.
Note that the intent of this demonstration is not to emphasize the utility of this batch size reduction method, but rather to show the degree to which training costs may vary with approach.
The practical consequence for assessing scalability is that training costs, and their scaling, are extremely sensitive to even minor perturbations in training protocol.
\\

\begin{figure*}[t!]
\includegraphics[width=\linewidth]{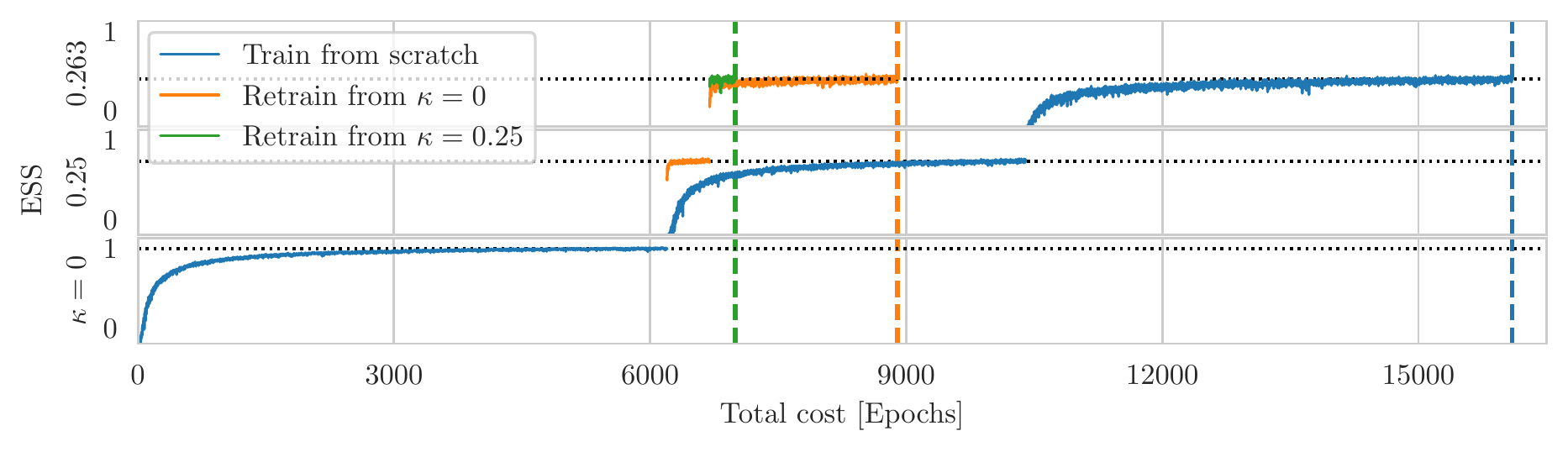}
\caption{
    Illustration of the reduction of training costs that can be obtained using transfer learning
    to train models for the Schwinger model with $\beta=2.0$ and lattice geometry $L\times T=8\times8$ for three different $\kappa$ targets.
    The ESS is measured on $6144$ samples each epoch.
    Training curves are sequentialized to illustrate the total cost of training models for $\kappa=0, 0.25, 0.263$ in order. The vertical lines indicate the overall cost in each case: blue for training each model from a random initialization, orange for retraining from an initial $\kappa=0$ model, and green for sequentially retraining from $\kappa=0 \rightarrow 0.25 \rightarrow 0.263$.
    Training is stopped when the target ESS is achieved in average over 100 epochs. Horizontal dotted lines mark the target $\text{ESS} = 0.9, 0.7, 0.45$ for $\kappa=0, 0.25, 0.263$, respectively.
    Further details are provided in Appendix~\ref{sec:numerics-appendix}.
}
\label{fig:param-transfer}
\end{figure*}

{\noindent \bf 2. Transfer learning can mitigate training costs.}
Flow models can be retrained between theories with the same lattice geometry but different action parameters.
\Cref{fig:param-transfer} illustrates the utility of this technique---``parameter transfer''---in training models for three different sets of Schwinger model target parameters.
The large cost of training a model from a random initialization must always be paid once.
However, as demonstrated, models targeting further parameters may be retrained from this initial model at significantly reduced cost compared with from-scratch training.
Sequentially retraining along a trajectory through parameter space provides further improvement~\cite{Hackett:2021idh}, at the cost of serializing training for different parameters.

Certain flow architectures, such as those parametrized by convolutional neural networks, additionally permit ``volume transfer'': a model trained to generate configurations with one geometry may be used to generate configurations with another~\cite{Boyda:2020hsi}.
This makes it possible to perform the bulk of training for smaller volumes than the final target, and enables methods that are intractable for larger volumes, e.g.,~training with on-the-fly HMC generation~\cite{Hackett:2021idh} or using exact fermion determinants \cite{Albergo:2021bna,Albergo:2022qfi}.
In practice, we often find that retraining on the larger volume is unnecessary. 
For models trained on smaller volumes and applied to larger ones with no retraining, there is a generic cost scaling relationship with changing volume, detailed in Sec.~\ref{sec:volume-scaling} below.

The consequence of these observations is that, for the purposes of assessing scalability, the cost scaling properties of training from a random initialization are irrelevant if transfer learning is employed.
Instead, the relevant scaling properties are those of the cost of retraining each new model, which will depend intimately on the set of physical parameters that are of interest in a particular study, and the order in which models are retrained between them.
\\

\begin{figure}[t!]
    \includegraphics[width=\linewidth]{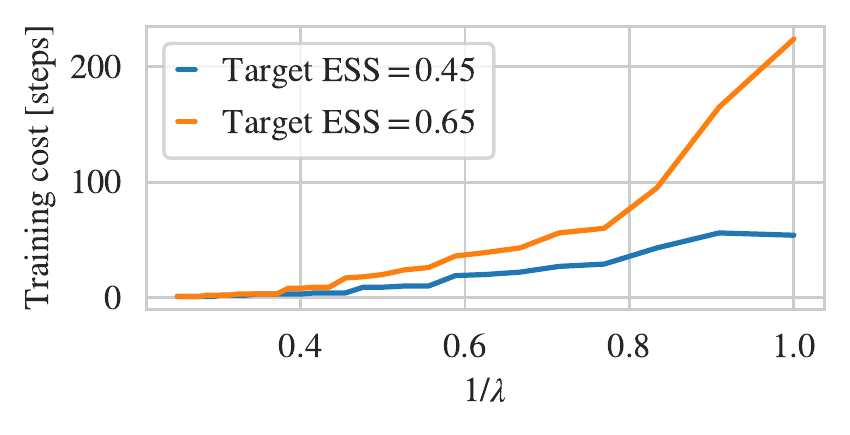}
    \caption{
    Demonstration in real scalar field theory, with $m^2 = -2$ on a $10 \times 10$ lattice, of the differences in training cost scaling behavior in $1/\lambda$ due to different choices of the stopping condition for training.
    Each model is retrained from $\lambda = 4$ to the target $\lambda$, halting training on the first step where the target ESS is achieved.
    The lack of smoothness is an inherent complication of quantifying scaling laws when using stopping conditions based on model quality, here due to the stochastic evaluation of the ESS well as the inherent noise of training.
    Further details are provided in Appendix~\ref{sec:numerics-appendix}.
    }
    \label{fig:training-stop-dependence}
\end{figure}

{\noindent \bf 3. Training cost scaling depends on training protocol.}\label{sec:training-cost-scaling}
Training costs and their scaling properties are determined by various choices made in training. As discussed in \cref{sec:stop-conditions}, this includes the stopping condition for training, without which cost scaling behavior is not well-defined. 
For example, training for a fixed number of iterations implies a fixed training cost but varying model quality. Alternatively, training to a target ESS results in fixed sampling costs while the training costs vary with the target theory.
The scaling of those training costs with the parameters of the target theory depends sensitively on the precise choice of target ESS, as illustrated in \cref{fig:training-stop-dependence}.
As already discussed in \cref{sec:stop-conditions}, this demonstration that scaling relations are highly specific to the training approach implies that, even if scaling behaviour can be determined empirically for some particular approach, it is unlikely to be generic across different training approaches, let alone across different theories or flow model architectures.

\subsection{Sampling costs}
\label{sec:sampling-costs}

In this section, we demonstrate various limitations of scaling laws in describing sampling costs in flow-based approaches.
In particular, as model quality determines sampling efficiency, the scaling of model quality with the parameters of the target theory is considered as a proxy for the scaling of sampling costs.
We provide arguments for and numerical illustrations of several key properties.
First, defining the maximum achievable model quality is a subtle question, since
\setlist{nolistsep}
\begin{enumerate}[noitemsep]
\item Achievable model quality is not a function of architecture alone.
\end{enumerate}
Moreover, the scaling of model quality with parameters of the theory is complex, since
\setlist{nolistsep}
\begin{enumerate}[resume, noitemsep]
\item Model quality scaling depends on training protocol, and
\item Different architectures scale differently.
\end{enumerate}
One scaling, however, can be derived generically:
\setlist{nolistsep}
\begin{enumerate}[resume, noitemsep]
\item For fixed models, quality scales exponentially in volume.
\end{enumerate}
Nevertheless, assessing scalability---i.e., utility at scale---from scaling laws is difficult, since
\setlist{nolistsep}
\begin{enumerate}[resume, noitemsep]
\item Best scaling does not imply best performance, and
    \item Architecture dependence implies theory dependence.
\end{enumerate}
Finally, we discuss the most relevant concern for scalability at present:
\begin{enumerate}[resume, noitemsep]
\item
Improving performance requires physics-informed algorithm design.
\end{enumerate}
While the provided numerical demonstrations exclusively use the direct sampling approach, the conclusions apply more generally to sampling algorithms including flow models or other machine-learned components.
\\

\begin{figure}
    \centering
    \includegraphics[width=\linewidth]{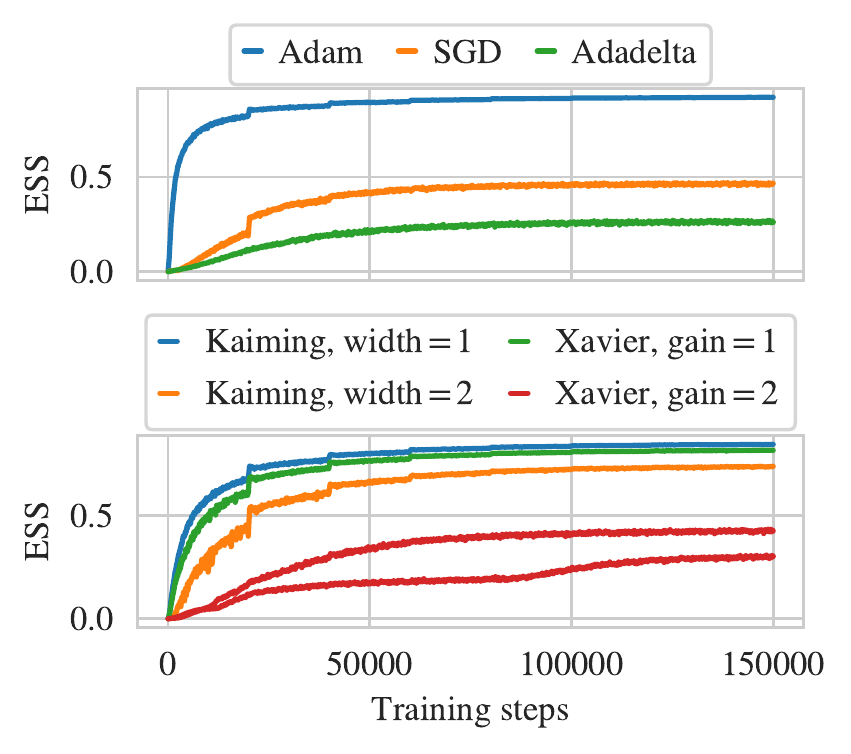}
    \caption{
        Example training curves demonstrating sensitivity of training dynamics and final model quality to the choice of optimization algorithm (top panel) and the distribution used to randomly initialize the model weights (bottom panel), for real scalar field theory on a $10 \times 10$ lattice with $m^2=-2$ and $\lambda=1$ (top), and $m^2=-4$ and $\lambda=1.5$ (bottom).
        Other than the optimizer or weights initialization, architectures and training protocols are the same for each curve.
        The legend for the top panel refers to the Adam optimizer of Ref.~\cite{kingma2017adam}, the Adadelta optimizer of Ref.~\cite{zeiler2012adadelta}, and stochastic gradient descent (SGD).
        In the legend for the bottom panel, Kaiming refers to the initialization procedure of Ref.~\cite{He:2015dtg}, the Pytorch 1.10 default, with ``width'' an overall rescaling of the distribution.
        Xavier refers to the procedure of Ref.~\cite{glorot2010understanding}, with ``gain'' a parameter of the method.
        The two red lines indicate examples for two different pseudorandom seeds; all other curves vary only at the $\sim 5-10\%$ level for different seeds.
        The learning rate is decayed by a factor of 2 every 20000 steps.
        The ESS is smoothed using a rolling window of width 250 steps.
        Further details are provided in Appendix~\ref{sec:numerics-appendix}.
    }
    \label{fig:no-saturation}
\end{figure}

{\noindent \bf 1. Achievable model quality is not a function of architecture alone.}
It is typical in gradient-descent-based optimization that after a period of rapid initial learning, optimization will enter a regime where model quality plateaus or improves only very slowly with further training.
It is tempting to interpret the resulting model as fully saturated in quality, i.e.~independent of training scheme and a function of architecture and target theory alone.
However, this is not the case.
As shown in \cref{fig:no-saturation}, changing only the precise choice of optimization algorithm or initial distribution of model weights can result in significantly different final ESS.
Further, training dynamics may be sensitive even to the pseudorandom seed used to generate the initial model parameters and training data.
The appearance of saturation may also be misleading, as demonstrated by the ``second-wind'' training dynamics in one of the examples in \cref{fig:no-saturation}, where an apparent plateau is broken by a second period of rapid learning with no associated change in training protocol.
Although a finite flow has finite expressivity, the results of any particular training procedure can only bound the capabilities of an architecture.
Practically, this implies that there is no notion of a ``fully trained'' model, and furthermore that model quality scaling relations cannot be quantified for an architecture class independent of training effects.
\\

\begin{figure}[t!]
    \includegraphics[width=\linewidth]{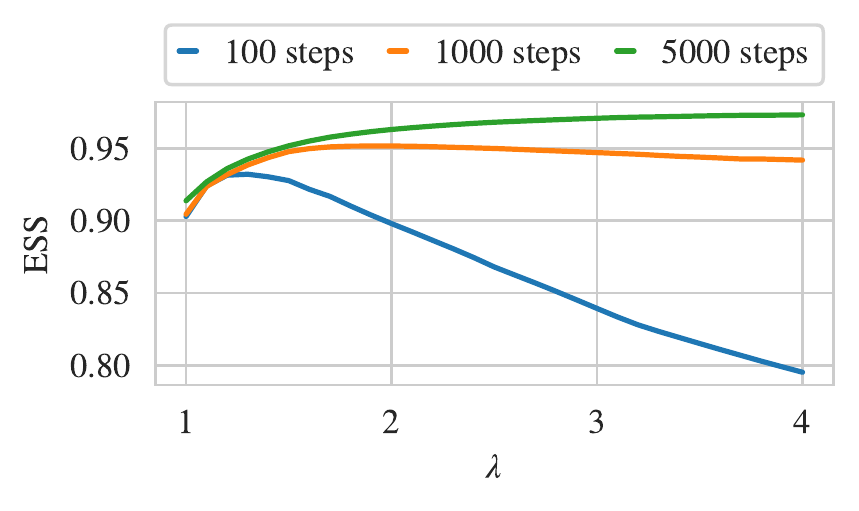}
    \caption{
    Demonstration in the context of real scalar field theory of how model quality as a function of $\lambda$ depends on the stopping condition used for training, for $m^2 = -2$ on a $10 \times 10$ lattice.
    Models are retrained directly from $\lambda = 1$ to the target $\lambda$, halting training after a fixed number of steps.
    Further details are provided in Appendix~\ref{sec:numerics-appendix}.
    }
    \label{fig:ess-stop-dependence}
\end{figure}

{\noindent \bf 2. Model quality scaling depends on training protocol.}
\label{sec:model-cost-scaling}
The scaling behavior of model quality is necessarily determined by the choice of training scheme, which governs how model weights vary with the parameters of the target theory. 
As for training costs, this includes the stopping condition for training.
\cref{fig:ess-stop-dependence} demonstrates how different stopping criteria---specifically, different choices of fixed expenditure on training---induce different functional forms in model quality as a function of target theory parameters when retraining between target parameters.
In this figure, models for all parameters are directly retrained from a single model; sequentially retraining instead would result in different functional forms again.
The practical consequences are that cost scaling laws are not unique even for a particular architecture and training method.
\\

\begin{figure}[t!]
    \includegraphics[width=\linewidth]{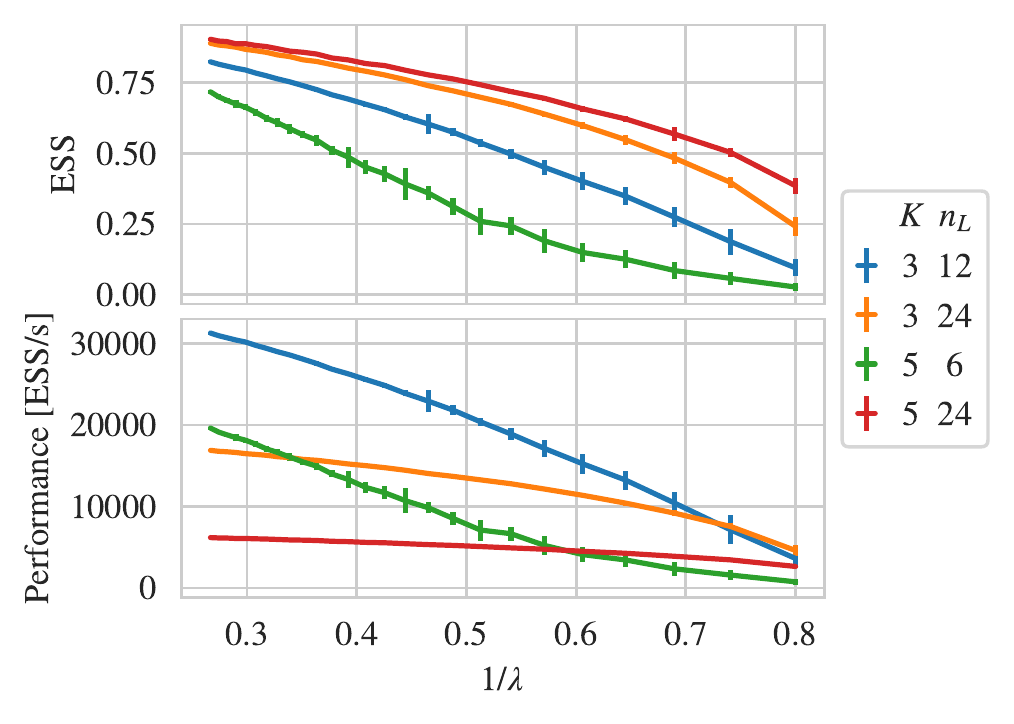}
    \caption{
    Scaling behavior of model quality with respect to theory parameters, for different model architectures applied to real scalar field theory with $m^2=-4$ on a $16 \times 16$ lattice.
    Each line denotes the ESS (top panel) and sampling performance
    (bottom panel) of a different affine coupling architecture, all trained using the same protocol.
    Performance is measured in effective samples per second, quantified for each model as its ESS divided by the computational cost of generating a configuration in units of RTX2080~Ti GPU-seconds. 
    Models are trained as described in Appendix~\ref{sec:numerics-appendix}.
    In the legend, $K$ denotes the convolutional kernel size and $n_L$ denotes the number of coupling layers.
    Further details are provided in Appendix~\ref{sec:numerics-appendix}.
    }
    \label{fig:xi-scaling-arch}
\end{figure}

{\noindent \bf 3. Different architectures scale differently.}
\Cref{fig:xi-scaling-arch} provides a simple demonstration of how the model quality of even very similar architectures may exhibit different scaling behaviors with the parameters of the target theory.
In this example, larger models (with more free parameters) exhibit better overall ESS and better scaling properties towards criticality---although the changing trade-off between ESS and $\Delta C_\text{samp}$ results in crossovers of the sampling performance, as discussed further below. The smooth dependence of ESS (and consequently performance) on target parameters implies useful extrapolation can be possible over a small range of target parameters for a fixed architecture, although we emphasize that these functional forms are necessarily specific to how the models are trained.
\\

\begin{figure}[t!]
    \includegraphics[width=\linewidth]{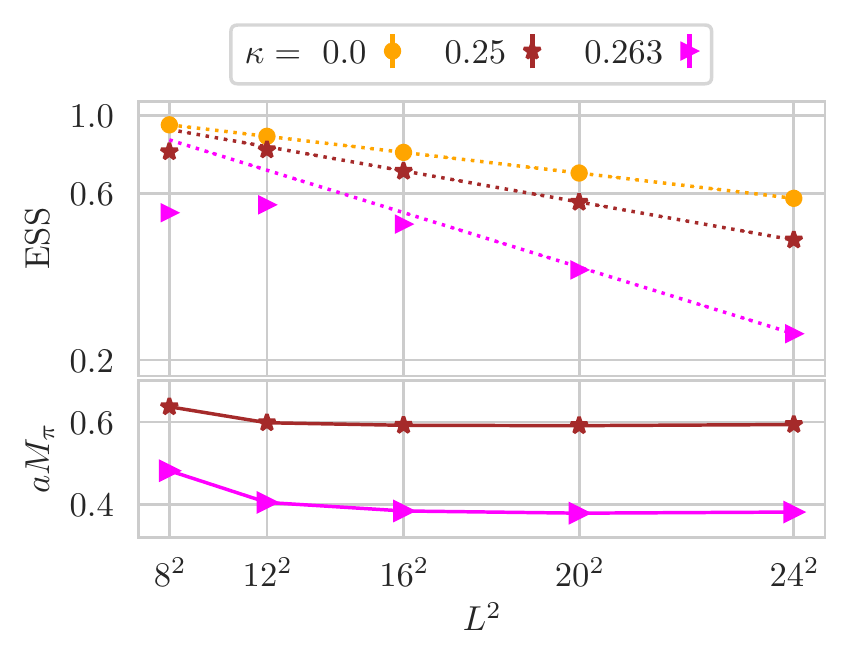}
    \caption{
        Volume scaling of the ESS for models trained to sample the Schwinger model at $\beta=2$ and three different values of the fermion mass, including the pure-gauge limit, $\kappa=0$. The top and bottom panels illustrate the volume dependence of the ESS and pseudoscalar mass $a M_\pi$, respectively.
        For each target $\kappa$, models are trained from scratch at $L=8$ and volume transferred without retraining.
        In the top panel, the dotted lines are extrapolations backwards from the largest volume using \cref{eq:volume-scaling}.
        Deviations of the markers below these lines correspond to better-than-exponential scaling at smaller volumes, where the lower panel shows significant finite-size effects in $a M_\pi$.
        The ESS is evaluated on $10^4$ samples at each volume.
        Error bars are smaller than the markers.
        Further details are provided in Appendix~\ref{sec:numerics-appendix}.
    }
    \label{fig:schwinger-volume-scaling}
\end{figure}

{\noindent \bf 4. For fixed models, quality scales exponentially in volume.}
\label{sec:volume-scaling}
As demonstrated throughout this work, generic cost scaling laws for flow models are difficult to obtain due to the complexity of the method. 
However, a particular restriction allows one to be derived analytically.
Specifically, a straightforward argument implies that the quality of a \emph{fixed} model constructed from a local\footnote{I.e., for which variables are transformed conditioned only on information in a local neighborhood, such as for flows parametrized by convolutional neural networks.} flow transformation degrades exponentially under volume transfer: if both $p$ and $q$ are defined for arbitrary volumes $V$ and may be characterized by correlation lengths $\xi_p$ and $\xi_q$, then when $L \gg \xi_p, \xi_q$ the integral defining the ESS factorizes over decorrelated subvolumes and scales as 
\begin{equation}
    \mathrm{ESS}(V) = \mathrm{ESS}(V_0)^{V/V_0} ~ .
    \label{eq:volume-scaling}
\end{equation}
\Cref{fig:schwinger-volume-scaling} demonstrates the onset of this effect in the large-volume regime for the Schwinger model, but we observe it to hold for other target theories as well.
This effect was first described in Ref.~\cite{Finkenrath:2022ogg}.

It is important to emphasize the limited applicability of this 
scaling relation.
Specifically, it applies only when the \emph{physical} extent $L/\xi_p$ is large, compared with the extent in lattice sites $L/a$.
Thus, while this relation may present an obstacle to flow-based sampling in the thermodynamic limit $L/\xi_p \rightarrow \infty$, it does not obstruct sampling in the continuum limit $\xi_p/a \rightarrow \infty$ with $L/\xi_p$ fixed.
Away from the thermodynamic limit, the breakdown of this relation can have counterintuitive effects: as shown in Fig.~\ref{fig:schwinger-volume-scaling}, we have observed better-than-exponential scaling in the Schwinger model when $L/\xi_p$ is small.
The exponential scaling relation furthermore does not apply between theories in different dimensions or with different numbers of internal degrees of freedom.
Most importantly, however, it does not relate different approaches or otherwise constrain what overall model quality (and thus quality of volume scaling properties) is achievable.
Finally, the argument for this scaling relation holds only for fixed models transferred without retraining.

This effect does not present any principled obstacle to flow-based sampling at state-of-the-art QCD parameters where typically $L/\xi_p = M_\pi L \sim 4-10$.
In practice, we are not interested in the thermodynamic limit, only control over finite-volume effects.
However, it does emphasize the importance of developing high-quality models to reach volumes of interest.
Sampling approaches which require modeling only subvolumes~\cite{Finkenrath:2022ogg} may control the effect directly.
\\

{\noindent \bf 5. Best scaling does not imply best performance.}
Of course, scaling of model quality is not the sole factor determining sampling efficiency.
In fact, in the demonstration of \cref{fig:xi-scaling-arch}, an architecture with one of the least favorable scaling behaviors provides the best performance.
For this small model, the lower ESS is more than compensated for by its low cost of evaluation, $\Delta C_\text{samp}$.
This emphasizes the important point that, ultimately, only sampling performance at parameters of interest matters; scaling laws are only important insofar as they can be used to predict the performance of an algorithm at one target set of parameters from its performance elsewhere.
\\

{\noindent \bf 6. Architecture dependence implies theory dependence.}
As demonstrated in \cref{fig:xi-scaling-arch}, even different choices of hyperparameters within an architecture class can lead to significantly different scaling of model quality with the parameters of the target theory.
As discussed in \cref{sec:flow-review}, much more significant differences in architecture are possible, from which we should expect an even greater variety in scaling behaviors.
Importantly, treating different physical theories \emph{requires} structurally different architectures.
For example, $\SU(N)$ variables cannot be flowed using the same transformations applicable to real scalar fields.
Further theory-specific engineering is necessary to encode symmetries and other physical features.
Given the dissimilarity between architectures required for different theories, it should not be assumed that features of model quality obtained for one theory apply for any other theory of interest; that is, studying the scaling of flow model sampling for toy models such as $\phi^4$-theory or even the Schwinger model provides little information about scaling properties for any flow-based approach to sampling gauge field configurations for QCD.
\\

\begin{figure}
    \centering
    \includegraphics[width=\linewidth]{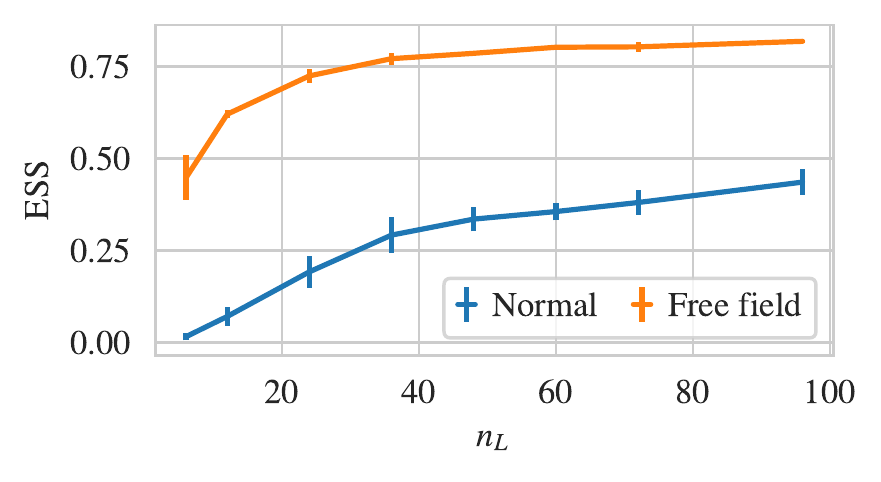}
    \caption{
        Dependence of ESS on number of coupling layers $n_L$ for two different architectures, with all other architecture and training hyperparameters fixed, for real scalar field theory on a $16 \times 16$ lattice with $m^2=-4$ and $\lambda=1.25$.
        The architectures differ by only the base distribution: independent normal on each site (``Normal''), and the free-field distribution with a learned pole mass (``Free field''; as described in Appendix~\ref{sec:numerics-appendix}).
        Each model is trained from a random initialization for 150k steps, which is sufficient to reach a slowly improving regime in all cases.
        Training uses batch size 1024, and the learning rate is decayed by a factor 2 every 10000 steps.
        Uncertainties for each point include the spread between two repeated experiments with different pseudorandom seeds.
        Further details are provided in Appendix~\ref{sec:numerics-appendix}.
    }
    \label{fig:hyperparam-sat}
\end{figure}

{\noindent \bf 7. Improving performance requires physics-informed algorithm design.}
As seen in Figs.~\ref{fig:phi4_bs_vs_lr_sched}, \ref{fig:param-transfer}, and \ref{fig:no-saturation},
training with a fixed protocol eventually enters a slowly improving regime, after which point the expense of additional training will not reliably provide a practical increase in sampling efficiency.
Similarly, as illustrated in \cref{fig:hyperparam-sat}, increasing the model size within fixed architecture classes often provides diminishing improvements in the model quality achieved by a fixed training protocol.
Although further training or increase in model size may provide sudden improvements (e.g.~second-wind training dynamics as in \cref{fig:no-saturation}), these saturation-like behaviors present a practical obstacle to increasing sampling efficiency using ``brute force'' application of additional computational power.

Instead, improving sampler efficiency requires more qualitative algorithmic improvements.
Among the infinite possible variations of model architecture, it is natural to expect that choices which incorporate a priori physics understanding will lead to better model quality.
This is illustrated in \cref{fig:hyperparam-sat}, which compares two closely related architectures for scalar field theory. Here it is clearly visible that the physics-informed modification is far more successful at improving model quality than simply increasing the model size.
Along the same lines, developing efficient flow-based sampling algorithms for QCD will require developing and testing new architectures, training schemes, and sampling approaches which incorporate a priori physics knowledge.

\section{Outlook}
\label{sec:outlook}

For flow-based approaches to the sampling of lattice field configurations, even small differences in the ML approach---architecture, training scheme, and how they are varied with the parameters of the target theory---result in not only different overall costs, but also different cost scalings with the parameters of the target theory.
In effect, each ML approach defines a different sampling algorithm with different cost scaling properties.
Further, because theory-specific modeling is required, scaling properties assessed in one theory should not be expected to generalize to others.
Taken together, this implies a very different paradigm for assessing algorithm scalability than has applied for QCD algorithms thus far, where scaling properties assessed in toy theories often translate directly to QCD.
For flow-based methods, assessing scalability will require direct, experimental investigation of applications to QCD itself, which has only just begun~\cite{Abbott:2022hkm}.

It is not yet clear what at-scale applications of flow-based sampling methods to QCD will look like, but we can speculate, and some key aspects are already clear.
For example, transfer learning and retraining will play a central role in mitigating training costs.
Thus, architectures well-suited for transfer learning across wide ranges of parameters and volumes will be important in order to exploit this technique.
Exponential volume scaling suggests that high-quality models will be necessary to achieve efficient sampling at scale. 
Coupled with the large overall scale of training costs, this suggests a paradigm of use very different from HMC.
As seen for large ML models in industry applications, such as GPT~\cite{brown2020language} and MT-NLG~\cite{smith2022using}, the typical paradigm is that significant computational resources and human time are invested over years in exploration and training. The resulting models can then be shared as a community resource, amortizing the bulk of training and development costs across the community as a whole.

Direct sampling with flow models has significant natural advantages over serial algorithms such as HMC, but requires high-quality models for efficient sampling.
However, there are many other sampling approaches where flows may be employed.
``Hybrid'' approaches involving both flow components and more traditional sampling algorithms can exploit lower-quality models while leveraging the decades of engineering invested in algorithms like HMC.
It is likely that the first at-scale application will involve such a hybrid approach.
Over the longer term, improving flow model technology will enable increasingly efficient sampling approaches.

While the growing body of work on flow-based sampling methods continues to provide promising early results, we have only just begun to explore the space of what is possible.
The broader field of machine learning is advancing rapidly, and experience dictates that we cannot anticipate what capabilities will be enabled by new developments.
Creativity guided by physical intuition remains our most effective means of making progress.

\backmatter

\bmhead{Acknowledgments}

The authors thank Heiko Strathmann for detailed comments on a draft of this manuscript, and Gurtej Kanwar for useful discussions. RA, DCH, FRL, PES, and JMU are supported in part by the U.S.\ Department of Energy, Office of Science, Office of Nuclear Physics, under grant Contract Number DE-SC0011090. PES is additionally supported by the National Science Foundation under EAGER grant 2035015, by the U.S.\ DOE Early Career Award DE-SC0021006, by a NEC research award, and by the Carl G and Shirley Sontheimer Research Fund. KC and MSA are supported by the National Science Foundation under the award PHY-2141336. MSA thanks the Flatiron Institute for their hospitality. DB is supported by the Argonne Leadership Computing Facility, which is a U.S. Department of Energy Office of Science User Facility operated under contract DE-AC02-06CH11357. This work is funded by the U.S.\ National Science Foundation under Cooperative Agreement PHY-2019786 (The NSF AI Institute for Artificial Intelligence and Fundamental Interactions, \url{http://iaifi.org/}). This work is associated with an ALCF Aurora Early Science Program project, and used resources of the Argonne Leadership Computing Facility, which is a DOE Office of Science User Facility supported under Contract DEAC02-06CH11357. The authors acknowledge the MIT SuperCloud and Lincoln Laboratory Supercomputing Center~\cite{reuther2018interactive} for providing HPC resources that have contributed to the research results reported within this paper. Numerical experiments and data analysis used PyTorch~\cite{NEURIPS2019_9015}, JAX~\cite{jax2018github}, Haiku~\cite{haiku2020github}, Horovod~\cite{sergeev2018horovod}, NumPy~\cite{harris2020array}, and SciPy~\cite{2020SciPy-NMeth}. Figures were produced using matplotlib~\cite{Hunter:2007}.

\begin{appendices}

\section{Details of numerical examples}
\label{sec:numerics-appendix}

Numerical illustrations of flow models trained to sample scalar field configurations are optimized to model the action
\begin{equation}
\begin{aligned}
    S(\phi) = \sum_x \Bigr[- & \sum_{\mu=1}^{d} \phi(x)\phi(x+\hat{\mu}) \\
    + \frac{1}{2} (m^2 &+ 2d)\phi(x)^2 + \lambda\phi(x)^4\Bigr] ~
\end{aligned}
\label{eqn:phi4-action}
\end{equation}
where $d=2$.
Architectures are stacks of $n_L$ affine coupling transformations with checkerboard variable partitioning, parametrized by convolutional neural networks, as implemented in Ref.~\cite{Albergo:2021vyo}.
All architectures have $n_L=24$ coupling layers, convolutional kernel size $K=5$, and two layers of hidden channels of width 12 within each neural network, except for the model used to generate the results shown in \cref{fig:xi-scaling-arch} where $n_L$ and $K$ differ as noted.
Reverse KL self-training protocols are also similar to Ref.~\cite{Albergo:2021vyo}, but with the addition of learning rate or batch size scheduling where indicated. Gradient clipping has also been applied, specifically rescaling all gradients by a common factor as necessary to prevent the norm over all gradients from exceeding 100.
Training and ESS evaluation both use batch size 16384, except as noted in \cref{fig:phi4_bs_vs_lr_sched}.
Results shown in \cref{fig:xi-scaling-arch} are initially trained for 150000 steps at $\lambda=1.25$, then for 500 steps at each subsequent $\lambda$, one after the other.
This is sufficient to reach a slowly improving regime for all $\lambda$, and produces comparable final results as from-scratch training or starting instead from $\lambda=4$.
For the results shown in Figs.~\ref{fig:training-stop-dependence}, \ref{fig:ess-stop-dependence}, and \ref{fig:xi-scaling-arch},
when retraining, the learning rate is set to $10^{-4}$ and the optimizer state is never reset (i.e.~it is carried over from training for the previous parameters).

Sampling from the ``Free field'' base distribution used in the demonstration of \cref{fig:hyperparam-sat} is accomplished using a layer which transforms the scalar field variables in momentum space.
We first perform a change of basis into Fourier space via $\phi(k) = \sum_x \phi(x) e^{-i k \cdot x}$.
We then transform the resulting momentum-space variables as $\phi'(k) = \sigma_k \phi(k)$, with
\begin{equation}
    \sigma_k^2 = \left( \mu + 2d - 2 \sum_{\nu = 1}^4 \cos(k_\nu) \right)^{-1}
\end{equation}
where $\mu$ is a learned parameter.
Finally, the updated variables are transformed back to position space via $\phi'(x) = \sum_k \phi'(k) e^{i k \cdot x} / \Omega$.
Comparing with the free-field (i.e.~$\lambda=0$) action in momentum space,
\begin{equation}
    S = \frac{1}{2} \sum_k \left( m^2 + 2d - 2 \sum_\nu \cos(k_\nu) \right) |\phi(k)|^2 ~ ,
\end{equation}
we see that if $\phi(k)$ are independent Gaussians for each $k$, then the transformation amounts to sampling from the free theory with learnable pole mass $\mu$.
This holds when $\phi(x)$ are independent Gaussians on each site, i.e.~if the layer is applied directly after the draw from the typical base distribution, as employed above.

Results shown in Figs.~\ref{fig:param-transfer} and \ref{fig:schwinger-volume-scaling} are obtained for models trained to sample from the theory defined by the two-flavor Wilson fermion action as employed in Ref.~\cite{Albergo:2022qfi}, specifically
\begin{equation}
    S_E(U) = -\beta \sum_{x} \mathrm{Re}{P(x)} - \log \det D[U]^\dag D[U]
\end{equation}
where $\beta = 2 / g_0^2$ encodes the bare gauge coupling $g_0$, and
\begin{equation}
    P(x) = U_{0}(x) U_{1}(x+\hat{0}) U^\dagger_{0}(x+\hat{1}) U^\dagger_{1}(x)
\end{equation}
is the plaquette.
The Wilson discretization of the Dirac operator is
\begin{equation}\begin{aligned}
    D[U](y, x) &= \delta(y-x)  \\
    - ~ \kappa \sum_{\mu=0,1}& \, \Big[  (1-\sigma_{\mu}) U_{\mu}(y) \delta(y - x  + \hat{\mu})  \\ &
    +  (1+\sigma_{\mu}) U^\dagger_{\mu}(y-\hat{\mu}) \delta(y - x - \hat{\mu}) \Big] \ ,
    \end{aligned}
\end{equation}
where $\sigma_0, \sigma_1 = \sigma_x, \sigma_y$ are the Pauli matrices and $\kappa = 1/(4+2 m_0)$ encodes the bare fermion mass $m_0$.
Architectures are similar to those of Ref.~\cite{Albergo:2022qfi}, with three differences: 1) models are built of 24 gauge-equivariant coupling layers rather than 48; 2) neural networks use standard convolutions with kernel size 3 rather than dilated convolutions; and 3) intermediate layers have 32 channels rather than 64.
All models used to create the results in \cref{fig:param-transfer,fig:schwinger-volume-scaling} have the same architecture.
Note that training and sampling uses exact computation of the fermion determinant, without any stochastic pseudofermion estimators like in Ref.~\cite{Abbott:2022zhs}.

To generate the results shown in \cref{fig:schwinger-volume-scaling}, models are self-trained with batch size 6144, for $\sim$~17k, 43k, and 85k steps for $\beta=0$, $0.25$, and $0.263$, respectively, at which points training has entered a slowly improving regime.
The initial learning rate $3 \times 10^{-4}$ is decayed by a factor 0.5 every 10k steps.
The pseudoscalar mass $M_\pi$ is obtained as the effective mass at the center of the lattice, as measured on configurations generated 
using flow-model proposals with independence Metropolis.

Throughout this work, the ESS metric used to evaluate model quality is estimated as
\begin{equation}
\text{ESS} = \frac{1}{\braket{w^2}_q} \approx \frac{ \left( 1/N \sum_i \tilde{w}_i \right)^2 }{ 1/N \sum_i \tilde{w}_i^2 } 
\end{equation}
where $\tilde{w}_i = \exp[-S(\phi_i)] / q(\phi_i)$.
This estimator is known to be positively biased at finite sample size, especially in the presence of mode collapse, where undersampling of mismodeled regions may result in an apparently large ESS when the true value may be near zero~\cite{Hackett:2021idh}.
This can be diagnosed using validation data drawn from the target distribution, using the reweighted estimator~\cite{Hackett:2021idh}
\begin{equation}
\text{ESS} = \frac{1}{\braket{w}_p} \approx \frac{ 1 }{ \left( \frac{1}{N} \sum_i \frac{1}{\tilde{w}_i} \right) \left( 1/N \sum_i \tilde{w}_i \right) }.
\end{equation}
Using the target-sample estimator to spot-check the model-sample ESS estimates presented throughout this work reveals no significant discrepancies.

\end{appendices}

\bibliography{main}

\end{document}